\documentclass[epj]{webofc}
\usepackage[varg]{txfonts}   
\usepackage{graphicx,color} 
\usepackage{bm}       
\usepackage{amsmath}  
\usepackage{amssymb}
\woctitle{21st International Conference on Few-Body Problems in Physics}
\begin{document}
\title{Two neutron decay of $^{16}$Be}
\author{A.E. Lovell\inst{1}\fnsep\thanks{\email{lovell@nscl.msu.edu}} \and
        F.M. Nunes\inst{1}\fnsep\thanks{\email{nunes@nscl.msu.edu}} \and
      I.J. Thompson\inst{2}\fnsep\thanks{\email{IJT@ianthompson.org}}
}

\institute{Department of Physics and Astronomy and National Superconducting Cyclotron Laboratory, Michigan State University, East Lansing, MI 48824, USA \and
Lawrence Livermore National Laboratory, L-414, Livermore, California 94551, USA
          }

\abstract{
Recently, the first example of two-neutron decay from the ground state of an unbound nucleus, $^{16}$Be, was seen (A. Spyrou, \emph{et. al.}, Phys. Rev. Lett. \textbf{108} 102501 (2012)).  Three-body methods are ideal for exactly treating the degrees of freedom important for these decays.   Using a basis expansion over hyperspherical harmonics and the hyperspherical R-matrix method, we construct a realistic model of $^{16}$Be in order to investigate its decay mode and the role of the two-neutron interaction.  The neutron-$^{14}$Be interaction is constrained using shell model predictions.  We obtain a ground state for $^{16}$Be that is under-bound by approximately 0.7 MeV with a width of approximately 0.17 MeV.  For such a system, an attractive three-body force must be included to reproduce the experimental ground state energy.
}
\maketitle
\section{Introduction}
\label{intro}
\indent Exotic nuclei can be found across the nuclear chart, from light two-proton and two-neutron halo nuclei, such as $^{17}$Ne and $^6$He, to heavy two-proton emitters, such as $^{45}$Fe, $^{48}$Ni, and $^{54}$Zn.  These nuclei can be modeled as three-body systems, a core plus two neutrons or protons, and three-body wave functions can be used to investigate their properties as well as to give insights into their decay modes.  (See for example \cite{RefGrig}.)  The difference between dinucleon decay (where two correlated neutrons or protons decay from a nucleus) and three-body decay (where two uncorrelated nucleons decay from a nucleus) is of particular interest.  \\
\indent Two-proton decay was first theorized in 1960 \cite{RefGold} and was then first experimentally verified over 40 years later in experiments such as \cite{RefJG}, \cite{RefMP}, and \cite{RefKM}.  These systems have been analyzed in terms of three-body models, where knowledge of their decay mechanism has led to a better understanding of their structure \cite{RefGrig}.  Although there has been much progress in the field of two-proton decay, much less is known about two-neutron radioactivity \cite{RefLVG}.  Recently, two-neutron decay from the continuum has been investigated theoretically \cite{RefLVGO26}.  \\
\indent An experiment performed in 2012 at the National Superconducting Cyclotron Laboratory first measured two-neutron radioactivity from the ground state of the unbound nucleus $^{16}$Be \cite{RefAS}.  Although $^{16}$Be is an ideal candidate for two-neutron radioactivity (the lowest experimentally measured state in $^{15}$Be is energetically inaccessible to one-neutron decay), there was controversy over this result when it was first published \cite{RefComment}.  In the original work \cite{RefAS}, the authors used two extreme models to explain the $^{16}$Be decay - one without a neutron-neutron interaction to model three-body decay [$^{16}$Be $\rightarrow$ $^{14}$Be + n + n] and one with two strongly correlated neutrons to model dineutron decay [$^{16}$Be $\rightarrow$ $^{14}$Be + (2n)].  However, in using a three-body formulation, we can exactly treat the degrees of freedom relevant for this decay - better understanding two-neutron decay and the neutron-neutron correlation in the nucleus.  

\section{Solutions to the three-body scattering problem}
\label{sec-2}

To solve the Schr\"odinger Equation in a three-body model, interactions between the three bodies must be defined.  Typically, experimentally measured levels in the core + n system (here, $^{15}$Be) would be used to constrain the interaction between these two bodies.  However, only one level in $^{15}$Be has been measured, at 1.8 MeV \cite{RefJS}, and therefore, shell model calculations are used to constrain this interaction.  Shell model calculations for $^{15}$Be \cite{RefBAB} give a $d_{5/2}$ state at 2.8 MeV, an $s_{1/2}$ state at 4.0 MeV, and a $d_{3/2}$ state at 7.0 MeV.  In order to match the experimentally measured state in $^{15}$Be, these shell model levels were lowered by 1 MeV in the current work.  In the three-body model, the depth of three Woods-Saxon potentials (one for each of the s,p, and d partial waves) is adjusted to reproduce the shell model levels.  \\
\indent The neutron-neutron interaction is taken to be the GPT interaction \cite{RefGPT}.  Although this interaction is simpler than others such as AV18 \cite{RefAV18} and Reid soft-core \cite{RefReid}, it reproduces NN scattering observables up to 300 MeV, which is more than sufficient for the energy region considered in this work.  The GPT interaction has often been used in the study of two-neutron halo nuclei (see for example, \cite{Ref12Be}, \cite{RefJPCS}, \cite{Ref11Li}, and \cite{Ref6He}).  A three-body interaction can also be included in order to adjust the final resonance energy to match the experimental value. \\
\indent For the three-body system, hyperspherical coordinates, $\rho$ and $\theta$, are used instead of the standard Jacobi coordinates \cite{RefB}.  This coordinate system allows the three-body wave function to be separated into hyperradial and hyperangular parts (where the hyperangular function can be constructed through a linear combination of known functions): 

\begin{equation}
\label{psi3b}
\Psi ^{JM} = \frac{1}{\rho ^{5/2}} \sum \limits _{K \gamma} ^{K_{max}} \chi ^{J} _{K \gamma} (\rho) \mathcal{Y} ^{JM} _{K \gamma} (\Omega _5, \sigma _1, \sigma _2, \xi).
\end{equation}

\noindent This expansion introduces the hypermomentum, $K$, related to the relative angular momenta, $l_x$ and $l_y$ \cite{RefB}.  $K_{\mathrm{max}}$ determines the size of the model space.  \\
\indent Then, all that must be solved is the system of coupled hyperradial equations

\begin{equation}
\left ( -\frac{\hbar ^2}{2m} \left [ \frac{d^2}{d\rho ^2} - \frac{(K+3/2)(K+5/2)}{\rho ^2}\right ] - E \right ) \chi ^J _{K \gamma} (\rho) + \sum \limits _{\gamma \prime K^\prime} V_{K K ^\prime \gamma \gamma ^\prime} (\rho) \chi ^J _{K ^\prime \gamma ^\prime} (\rho) = 0,
\label{radeq}
\end{equation}

\noindent with scattering boundary conditions

\begin{equation}
\chi ^{LKK_i} _{\gamma \gamma _i} (\kappa \rho) \rightarrow \frac{i}{2} \left [ \delta ^{KK_i} _{\gamma \gamma _i} H^- _{K+3/2} (0, \kappa \rho) - \textbf{S} ^{LKK_i} _{\gamma \gamma _i} H^+ _{K+3/2} (0, \kappa \rho) \right ],
\end{equation}

\noindent where $\kappa = \frac{\sqrt{2mE}}{\hbar}$.  The index $\gamma$ contains the quantum numbers $l$ (total orbital angular momentum of the two neutrons relative to the core), $S$ (total spin of the two neutrons), $j$ (total angular momentum of the two neutrons relative to the core), $I$ (the spin of the core), $l_x$ (relative orbital angular momentum of the two neutrons), and $l_y$ (relative orbital angular momentum of the core + (2n) system).  The subscript $i$ labels the incoming channel.  \\
\indent Equation (\ref{radeq}) is solved using the hyperspherical R-matrix method \cite{RefB}.  Through the R-matrix method, phase shifts can be calculated, allowing for the extraction of resonance energies and widths.  

\section{Preliminary Phase Shift Analysis and Future Work}
\label{sec-3}
\begin{figure}
\centering
\sidecaption
\includegraphics[width=8cm,clip]{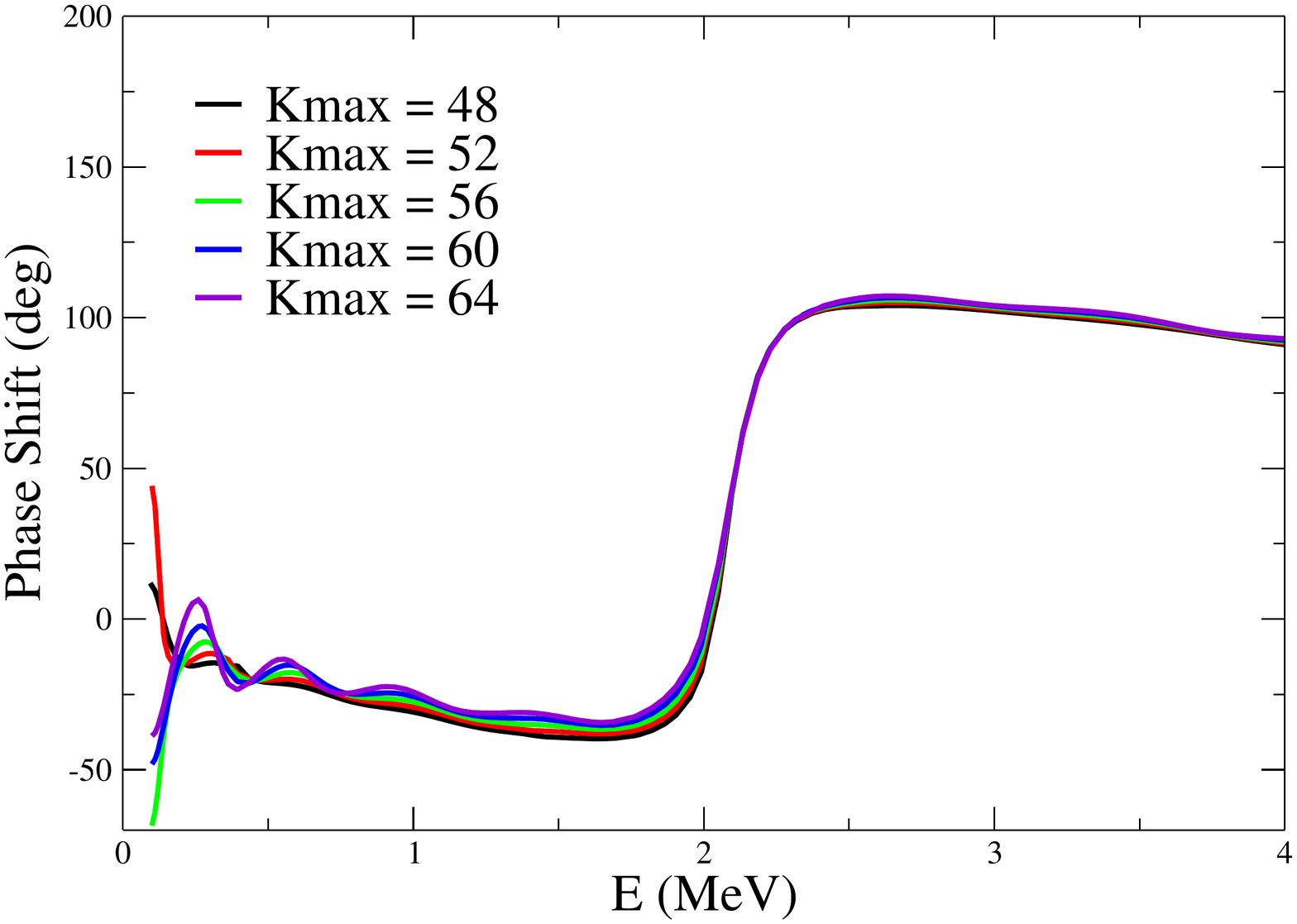}
\caption{Phase shifts as a function of three-body energy of the $K=0$ channel for the largest values of $K_{\mathrm{max}}$ used in the calculation.}
\label{fig-1}       
\end{figure}

In this work, energies are extracted from the phase shifts in the channel $K=0$ at the points where the energy derivative of the phase shift has a maximum.  In Figure \ref{fig-1}, we show the convergence of the these phase shifts.  This system required a large model space for convergence, including 80 hyperradial basis states, hypermomentum greater than 64, and over 80 hyperangular basis states.  For values greater than $K_{\mathrm{max}}=64$, numerical inaccuracies were induced due to the strong repulsive centrifugal barrier.  By finding the peak of the derivative of the phase shifts in Figure \ref{fig-1}, the resonance energy was calculated to be 2.07 MeV.  \\
\indent This ground state energy found above is under-bound by about 0.7 MeV when compared to the experimental ground state energy of 1.35 MeV \cite{RefAS}.  Traditionally, three-body forces in three-body models account for the additional binding coming from degrees of freedom that are excluded from the model space.  Here, an attractive three-body force must be added to the model in order to reproduce the experimental ground state of $^{16}$Be.  \\
\indent A rough analysis of the $K=0$ phase shift shows that the width is approximately 0.17 MeV.  Due to the coupled-channel nature of these calculations, the interpretation of the phase shifts is not straightforward.  Because of effects from the couplings between the channels, phase shifts do not rise sharply through $90 ^{\circ}$ at the resonance energy.  Therefore, it is not clear whether the ground state of the calculated system is given simply by the lowest resonance energy extracted from a single phase shift (as computed here) or if these phase shifts have to be weighted and combined, in the case of several resonance energies around the same region.  One such method for this weighting would be to calculate the widths of the state through an integral relationship, such as that used in \cite{RefNoll}.  A study comparing the various approaches is underway.\\


\begin{acknowledgement}
The authors would like to acknowledge iCER and the High Performance Computing Center at Michigan State University for their computational resources.  This work was supported by the National Science Foundation under Grant No. PHY-1403906.  This work was performed under the auspices of the U.S. Department of Energy through NNSA contract  DE-FG52-08NA28552 and by Lawrence Livermore National Laboratory under Contract DE-AC52-07NA27344.
\end{acknowledgement}

%

\begin{thebibliography}{99}
%
%
\bibitem{RefGrig}
L.V. Grigorenko and M.V. Zhukov, Phys. Rev. C \textbf{68} 054005 (2003)
\bibitem{RefGold}
V.I. Goldansky, Nucl. Phys. \textbf{19} 482 (1960)
\bibitem{RefJG}
J. Giovinazzo, \emph{et. al.}, Phys. Rev. Lett. \textbf{89} 102501 (2002)
\bibitem{RefMP}
M. Pf\"utzner, \emph{et. al.}, Eur. Phys. Jour. A \textbf{14} 279 (2002)
\bibitem{RefKM}
K. Miernik, \emph{et. al.}, Phys. Rev. Lett. \textbf{99} 192501 (2007)
\bibitem{RefLVG}
L.V. Grigorenko, \emph{et. al.}, Phys. Rev. C \textbf{84} 021303(R) (2011) 
\bibitem{RefLVGO26}
L.V. Grigorenko and M.V. Zhukov, Phys. Rev. C \textbf{91} 064617 (2015)
\bibitem{RefAS}
A. Spyrou, \emph{et. al.}, Phys. Rev. Lett. \textbf{108} 102501 (2012)
\bibitem{RefComment}
F.M. Marques, \emph{et. al.}, Phys. Rev. Lett. \textbf{109} 239201 (2012)
\bibitem{RefJS}
J. Snyder, \emph{et. al.}, Phys. Rev. C \textbf{88} 031303(R) (2013)
\bibitem{RefBAB}
B.A. Brown, \emph{Private Communications}, Nov. 21, 2013
\bibitem{RefGPT}
D. Gogny, P. Pires, and R. De Tourreil, Phys. Lett. \textbf{32B} 591 (1970)
\bibitem{RefAV18}
R.B. Wiringa, V.G.J. Stokes, and R. Schiavilla, Phys, Rev. C \textbf{51} 38 (1995)
\bibitem{RefReid}
Roderick V. Reid, Jr. Ann. Phys. \textbf{50} 411 (1968)
\bibitem{Ref12Be}
F.M. Nunes, J.A. Christley, I.J. Thompson, R.C. Johnson, and V.D. Efros, Nucl. Phys. A \textbf{609} 43 (1996)
\bibitem{RefJPCS}
N.K. Timofeyuk, I.J. Thompson, and J.A. Tostevin, J. Phys. Conference Series \textbf{111} 012034 (2009)
\bibitem{Ref11Li}
I. Brida, F.M. Nunes, and B.A. Brown, Nucl. Phys. A \textbf{775} 23 (2006)
\bibitem{Ref6He}
I. Brida and F.M. Nunes, Int. J. Mod. Phys. E \textbf{17} 2374 (2008)
\bibitem{RefB}
Ian J. Thompson and Filomena M. Nunes, \textit{Nuclear Reactions for Astrophysics} (Cambridge University Press, United Kingdom, 2009) 221-225,279-286
\bibitem{RefNoll}
K.M. Nollett, Phys. Rev. C \textbf{86} 044330 (2012)
\end{thebibliography}
%
%

\end{document}